\documentclass[prl,aps,twocolumn,superscriptaddress]{revtex4}
\usepackage{graphicx}
\usepackage{epstopdf}
\usepackage{latexsym}
\usepackage{amsmath}
\begin{document}

\title{Nodal quasiparticles and the onset of spin density wave order in the cuprates}

\author{Andrea Pelissetto}
\affiliation{Dipartimento di Fisica dell'Universit\`a di Roma ``La Sapienza'' and INFN,
P.le Aldo Moro 2, I-00185 Roma, Italy}

\author{Subir Sachdev}
\affiliation{Department of Physics, Harvard University, Cambridge,
MA 02138, USA}

\author{Ettore Vicari}
\affiliation{Dipartimento di Fisica dell'Universit\`a di Pisa and INFN,
Largo Pontecorvo 2, I-56127 Pisa, Italy}

\date{\today}

\begin{abstract}
  We present a theory for the onset of spin density wave order in the
  superconducting ground state of the cuprates. We compute the scaling
  dimensions of allowed perturbations of a `relativistic' fixed point with
  O(4)$\otimes$O(3) symmetry, including those associated with the fermionic
  nodal Bogoliubov quasiparticles. Analyses of up to six loops show that all
  perturbations with square lattice symmetry are likely irrelevant.  We
  demonstrate that the fermion spectral functions are primarily damped by the
  coupling to fluctuations of a composite field with Ising nematic order. We
  also discuss the influence of quenched disorder.
\end{abstract}
\maketitle

A large number of experiments have investigated the magnetic correlations in
the cuprate superconductors \cite{jtran}.  There is good evidence for
collinear spin density wave (SDW) correlations near the wavevectors
\begin{displaymath}
{\bf K}_1 =  
\left(\frac{2\pi}{a}\right)\left(\frac{1}{2}-\vartheta,\frac{1}{2}\right)~~,~~{\bf
K}_{2} = \left(\frac{2 \pi}{a}\right)
\left(\frac{1}{2},\frac{1}{2}-\vartheta\right),
\end{displaymath}
where $a$ is the square lattice spacing and the incommensurate $\vartheta$
varies with the hole doping concentration, $\delta$.  Especially in the La
based compounds, this order is best formed near $\delta =1/8$, where
$\vartheta \approx 1/8$. This SDW order coexists with superconductivity (SC),
and a theory \cite{demler} for the tuning of the transition between the SDW+SC
and SC states by an applied magnetic field has been tested in a number of
experiments \cite{katano,lake,jtran2,boris,mesot}.

In this paper, we will examine the interplay between the SDW order and the
fermionic Bogoliubov quasiparticle excitations of the superconductor. Because
of the $d$-wave nature of the Cooper pairs, these quasiparticles have the
spectrum of massless Dirac particles whose energy vanishes at four wavevectors
(the nodal points) in the Brillouin zone $(\pm Q, \pm Q)$; these nodal
wavevectors bear no relation to the SDW ordering wavevectors ${\bf K}_{1,2}$.
Using a theory for the critical fluctuations of the SDW order developed in
Refs.~\cite{demler,zachar,vicari1}, we will show that the dominant coupling
between nodal quasiparticles and the SDW order is to a composite Ising nematic
order \cite{nematic}, associated with the choice of the SDW ordering between
the ${\bf K}_1$ and ${\bf K}_2$ wavevectors. We emphasize that the unique selection of 
Ising nematic order here is not a phenomenological assumption, but a
non-trivial consequence of the internal structure of the SDW fixed point of De
Prato {\em et al.} \cite{vicari1}; indeed, this is the main point of this
paper.  Our theory predicts nodal quasiparticle spectral functions which have
been measured in numerous photoemission and scanning tunnelling microscopy
experiments.  The importance of the Ising nematic order in damping nodal
quasiparticles was noted recently by Kim {\em et al.\/} \cite{kim} in a
different model \cite{qptd}, and we will compare our results to theirs.

Our theory is expressed in terms of the complex-valued order parameters
$\Phi_{1 i}$ and $\Phi_{2 i}$, where $i =x,y,z$ denotes spin components,
which are related to the spin operator, $S_i$, at position ${\bf r}$ and
imaginary time $\tau$ by
\begin{equation}
S_i({\bf r},\tau)= {\rm Re}\left[e^{i {\bf K}_1 \cdot {\bf r}}\,
\Phi_{1i}({\bf r},\tau)+e^{i {\bf K}_2 \cdot {\bf r}}\,
\Phi_{2i}({\bf r},\tau)\right]. \label{SPhi}
\end{equation}
This parameterization implies the transformation properties of $\Phi_{1,2i}$ under the symmetry
operations on the underlying square lattice Hamiltonian, which are summarized in Fig~\ref{table}.
\begin{figure}
\begin{tabular}{c|ccccc}
& $T_x$ & $T_y$ & $~~~R~~~$ & $~~~I~~~~$ & $~~~\mathcal{T}~~~$\\
\hline
$\Phi_{1i}$ & $- e^{-i\vartheta} \Phi_{1i}$ & $- \Phi_{1i}$ & $\Phi_{2i}$ & $\Phi_{1i}^\ast$ & $- \Phi_{1i}$ \\
$\Phi_{2i}$ & $-  \Phi_{2i}$ & $- e^{-i\vartheta}\Phi_{2i}$ & $\Phi_{1i}^\ast$ & $\Phi_{2i}^\ast$ & $- \Phi_{2i}$ \\
$\Psi_{1 \alpha}$ & $e^{iQ} \Psi_{1\alpha}$ & $e^{iQ} \Psi_{1\alpha}$ &$ i \tau^z \Psi_{2\alpha}$ & $\Psi_{2\alpha}$ & $- \tau^y \Psi_{1\alpha}$\\
$\Psi_{2 \alpha}$ & $e^{-iQ} \Psi_{2\alpha}$ & $e^{iQ} \Psi_{2\alpha}$ &$ - i \varepsilon_{\alpha\beta} \left[ \Psi_{1\beta}^\dagger \tau^x \right]^T $ 
& $\Psi_{1\alpha}$ & $- \tau^y \Psi_{2\alpha}$\\
\end{tabular}
\vspace{0.1in}
\caption{
Transformations of the fields under operations which generate the symmetry group: $T_{x,y} = $ translation by a lattice
spacing in the $x,y$ directions, $R = $ rotation about a lattice site by 90$^\circ$, $I = $ reflection about the $y$ axis on a lattice site,
and $\mathcal{T} = $ time reversal. The theory is also invariant under spin rotations, with $i$ a vector index and $\alpha,\beta$ spinor indices.
We define $\mathcal{T}$ as an invariance of the imaginary time path integral, in which $\Phi_{1,2i}^\ast$ transform as the complex
conjugates of $\Phi_{1,2i}$, while $\Psi_{1,2\alpha}^\dagger$ are viewed as independent complex Grassman fields which transform
as $\Psi_{1,2\alpha}^\dagger \rightarrow \Psi_{1,2\alpha}^\dagger \tau^y$.
}
\label{table}
\end{figure}
By writing down all terms invariant under these operations, and expanding in
powers and gradients of $\Phi_{1,2i}$, we obtain the following quantum field
theory for the fluctuations of the SDW order \cite{demler,zachar} with action
$\int d^2 r d\tau \mathcal{L}_\Phi$ and
\begin{eqnarray}
&& \mathcal{L}_\Phi =\vert\partial_\tau
\Phi_{1}\vert^2 +v_1^2\vert\partial_x \Phi_{1}\vert^2
+v_2^2\vert\partial_y \Phi_{1}\vert^2  \nonumber\\ && +\vert\partial_\tau
\Phi_{2}\vert^2+  v_2^2\vert\partial_x
\Phi_{2}\vert^2 +v_1^2\vert\partial_y \Phi_{2}\vert^2
+r(\vert\Phi_{1}\vert^2+\vert\Phi_{2}\vert^2)\nonumber\\
&& +\frac{u_{1}}{2}(\vert\Phi_{1}\vert^4+\vert\Phi_{2}\vert^4)
+\frac{u_{2}}{2}(\vert\Phi_{1}^2\vert^2+\vert\Phi_{2}^2\vert^2)
\nonumber\\
&& 
  +w_{1}\vert\Phi_{1}\vert^2 \vert\Phi_{2}\vert^2
  +w_{2}\vert\Phi_{1} \cdot \Phi_{2}\vert^2 
  +w_{3}\vert\Phi_{1}^* \cdot \Phi_{2}\vert^2 ,
\label{lgwhor} 
\end{eqnarray}
where $v_{1,2}$ are spin-wave velocities, $r$ is the coupling which tunes the
system across the transition to a state with SDW order, and $u_{1,2}$ and
$w_{1,2,3}$ are the crucial quartic couplings which stabilize and select the
SDW order. Linear spatial derivative terms such as $ \Phi_1^\ast \partial_x
\Phi_1$ are also permitted, but can be absorbed by a redefinition of the
incommensuration $\vartheta$. For some of the cuprates with $\delta \approx
1/8$, the ordering is commensurate with $\vartheta=1/8$; in this case a
lock-in term $\sim (\Phi_{1}^2)^4 + (\Phi_{2}^2)^4$ is allowed, but this
eighth order term is clearly irrelevant near the critical point, and can be
safely neglected in our considerations here.

The weight of the experimental evidence, reviewed {\em e.g.\/} in
Ref.~\onlinecite{jtran}, is that the SDW order in the cuprates is collinear in
spin space, and picks a definite spatial direction by a choice of condensing
either $\Phi_1$ or $\Phi_2$. So a particular state has
\begin{equation}
\langle \Phi_{1i} \rangle = n_i e^{i\theta}~~,~~\langle \Phi_{2i} \rangle = 0,
\end{equation}
or an equivalent state with $1 \leftrightarrow 2$, with $n_i$ an arbitrary
real vector and $e^{i \theta}$ a common complex phase for the components of
$\Phi$. In mean field theory, such a state is the ground state of the theory
$\mathcal{S}_\Phi$ in a subset of the region of parameters satisfying $r<0$, $w_1 > 0$, $w_1 +
w_2 + w_3 > 0$, $u_2 < {\rm Min}[0,w_1-u_1,w_1 + w_2 + w_3 - u_1,(w_1 + w_2 - u_1)/2, (w_1 +
w_3 - u_1)/2]$.  Going beyond
mean field theory, a detailed analysis of the transition into this SDW state
under the Lagrangian $\mathcal{L}_\Phi$ was provided in
Ref.~\onlinecite{vicari1}.  It was found that such a transition was in the
domain of attraction of a fixed point with O(4)$\otimes$O(3) symmetry within
two three-dimensional perturbative schemes: ({\em i\/}) the massive
zero-momentum scheme (MZM) defined in the unbroken phase (to six loops), and
({\em ii\/}) the minimal subtraction scheme ($\overline{\rm MS}$) without expansion in
$3-d$ ($d$ is the spatial dimension, and $d=2$ was set after
renormalization) defined in the massless critical theory (to five loops).
High-order perturbative calculations indicate the stability of this point also
in the SDW model $\mathcal{L}_\Phi$, {\em i.e.\/} the quartic perturbations
present in ${\mathcal L}_{\Phi}$ turn out to be irrelevant at $O(4)\otimes
O(3)$ fixed point.  The values of the quartic couplings in the fixed-point
Lagrangian $\mathcal{L}_\Phi^\ast$ obey $w_1^\ast = u_1^\ast - u_2^\ast$ and
$w_2^\ast = w_3^\ast = u_2^\ast$, and the spin-wave velocities are equal
$v_1^\ast = v_2^\ast$. The symmetry of the fixed point becomes explicit by
introducing a 12-component real field $\varphi_{ai}$, with $a=1\ldots 4$,
related to the SDW order
\begin{equation}
\Phi_{1i} = \varphi_{1i} + i \varphi_{2i},~~~\Phi_{2i} = \varphi_{3i} + i \varphi_{4i};
\end{equation}
now the O(4) and O(3) rotations act on the $a$ and $i$ indices respectively.
The critical exponents
associated with this fixed point are the correlation length exponent, $\nu = 0.9 (2)$,
and the anomalous dimension of the SDW order parameter, $\eta = 0.15 (10)$.

We are now ready to turn to our new results on the coupling between the above
SDW fluctuations and the excitations of the superconductor.

Density fluctuations of the superconducting condensate will couple to $|\Phi_1|^2 + |\Phi_2|^2$. For
short-range repulsive interactions between the Cooper pairs, this coupling has scaling dimension \cite{frey}
$(2-3\nu)/2 < 0$ and so is irrelevant. The long-range Coulomb interactions further suppress density fluctuations,
and so we need not consider this coupling further.

Far more interesting, and subtle, are the couplings between $\Phi_{1,2}$ and the fermionic nodal quasiparticles. 
Let us denote the electron annihilation operator with momenta in the vicinity of the nodes as
$(Q,Q)$, $(-Q,Q)$, $(-Q,-Q)$, and $(Q, -Q)$ by $f_{1\alpha}$, $f_{2\alpha}$, $f_{3\alpha}$, and $f_{4 \alpha}$
respectively; here $\alpha = \uparrow, \downarrow$ is an electron spin index.
Next, we introduce the 4-component Nambu spinors 
\begin{equation}
\Psi_{1\alpha} =
\left( \begin{array}{c} f_{1\alpha} \\ \varepsilon_{\alpha\beta} f_{3\beta}^{\dagger} \end{array} 
\right)~~,~~\Psi_{2\alpha} =
\left( \begin{array}{c} f_{2\alpha} \\ \varepsilon_{\alpha\beta} f_{4\beta}^{\dagger} \end{array}
\right) 
\end{equation}
where
$\varepsilon_{\alpha\beta}=-\varepsilon_{\beta\alpha}$ and
$\varepsilon_{\uparrow \downarrow} = 1$. 
We will use Pauli matrices $\tau^i$ which act on the Nambu particle-hole 
space, while $\sigma^i_{\alpha\beta}$ will
act on spin space. In the vicinity of the nodal points, we can expand the standard Bogoliubov Hamiltonian
of a $d$-wave BCS superconductor in term of gradients of $\Psi_{1,2}$, and obtain the low energy theory
$\mathcal{S}_\Psi = \int d^2 r d \tau \mathcal{L}_\Psi$ with
\begin{eqnarray}
&& \mathcal{L}_{\Psi} = 
\Psi_{1}^{\dagger}  \left(
\partial_\tau -i \frac{v_F}{\sqrt{2}} (\partial_x + \partial_y) \tau^z  -i \frac{v_\Delta}{\sqrt{2}} (-\partial_x + 
\partial_y) \tau^x \right) \Psi_{1}   \nonumber \\
&&+  \Psi^\dagger_2 \left(
\partial_\tau - i \frac{v_F}{\sqrt{2}} (-\partial_x + \partial_y) \tau^z -i  \frac{v_\Delta}{\sqrt{2}} (\partial_x + \partial_y)  \tau^x \right) \Psi_{2} .\label{dsid1}
\end{eqnarray}
Here $v_{F}$ and $v_{\Delta}$ are the fermionic velocities normal and parallel to the 
underlying Fermi surface. As was the case with $\mathcal{S}_\Phi$, the transformations of the fermionic fields
$\Psi_{1,2}$ are also crucial to our analysis, and these are summarized in Fig.~\ref{table}.

We now write down the most general couplings between the SDW fields $\Phi_{1,2}$ and the Dirac
fermion excitations $\Psi_{1,2}$ which are allowed by the symmetries in Fig.~\ref{table}. The translational
symmetries $T_{x,y}$ immediately rule out any terms linear in $\Phi$: the SDW fluctuations
scatter the electrons by wavevectors $K_{1,2}$ and these do not, in general, connect 
the low energy excitations at any pair of nodal points \cite{qptd}. Moving on to terms 
quadratic in $\Phi$, which are generated by a virtual scattering process to a high energy intermediate
fermion excitation (see also Ref.~\onlinecite{berg}), all the allowed terms are
\begin{eqnarray}
&& \mathcal{L}_1 = \lambda_1 \left( |\Phi_{1} |^2 + |\Phi_{2}|^2 \right) 
\left( \Psi^\dagger_{1} \tau^z \Psi_{1} +  \Psi^\dagger_{2} \tau^z \Psi_{2} \right) \nonumber \\
&& \mathcal{L}_2 = \lambda_2 \left( |\Phi_{1} |^2 - |\Phi_{2}|^2 \right) 
\left( \Psi^\dagger_{1} \tau^x \Psi_{1} +  \Psi^\dagger_{2} \tau^x \Psi_{2} \right) \nonumber \\
&& \mathcal{L}_3 = \epsilon_{ijk} \Bigl[  \label{lambda} \\
&& \left(\Phi^\ast_{1j} \Phi_{1k} + 
\Phi_{2j}^\ast \Phi_{2k} \right) \left( - \lambda_3 \Psi^\dagger_{2}
\tau^x \sigma^i \Psi_{2} + \lambda_3^\prime \Psi^\dagger_{1}
\tau^z \sigma^i \Psi_{1} \right) \nonumber \\
&&+  \left(\Phi^\ast_{1j} \Phi_{1k} - 
\Phi_{2j}^\ast \Phi_{2k} \right) \left(  \lambda_3 \Psi^\dagger_{1}
\tau^x \sigma^i \Psi_{1} - \lambda_3^\prime \Psi^\dagger_{2}
\tau^z \sigma^i \Psi_{2} \right) \Bigr]. \nonumber
\end{eqnarray}

Key to our remaining analysis are the scaling dimensions of the $\lambda$
couplings at the O(4)$\otimes$O(3) fixed point $\mathcal{L}_\Phi^\ast$
discussed earlier, at which the fermions are decoupled from the SDW order. The
free Dirac fermion theory $\mathcal{L}_\Psi$ is also invariant under the same
scaling transformation as $\mathcal{L}_\Phi^\ast$, with $\mbox{dim}[\Psi] =
1$, and this allows us to obtain the $\mbox{dim}[\lambda]$ from a knowledge of
the dimensions of the associated composite operators of $\Phi$ under
$\mathcal{L}_\Phi^\ast$.  In particular, we perturb $\mathcal{L}_\Phi^\ast$ by
the same quadratic $\Phi$ operators above but without the fermion terms, and
subtract 2 from the scaling dimension of such a perturbation. So {\em e.g.\/}
we consider the theory $\mathcal{L}_\Phi^\ast + \widetilde{\lambda}_1 \left(
  |\Phi_{1} |^2 + |\Phi_{2}|^2 \right)$ and obtain $\mbox{dim}[\lambda_1] =
\mbox{dim}[\widetilde{\lambda}_1] -2$.  We classify such perturbations under
representations of O(4)$\otimes$O(3), with ${\bf D}_K$ indicating a dimension
${\bf D}$ representation of O($K$). Then $\widetilde{\lambda}_1$ corresponds
to a $({\bf 1}_4, {\bf 1}_3)$ operator, $\widetilde{\lambda}_2$ to $({\bf
  9}_4, {\bf 1}_3)$, and $\widetilde{\lambda}_3$ to $({\bf 6}_4,{\bf 3}_3)$.

The $({\bf 1}_4, {\bf 1}_3)$ operator tunes away from the transition by
changing $r$, and so
\begin{equation}
\mbox{dim}[\lambda_1] = 1/\nu - 2. \label{dim1}
\end{equation}
From the value of $\nu$ quoted earlier, we conclude that $\lambda_1$ is safely
irrelevant.

The second coupling between $\Phi$ and the fermions involves the operator
$\phi \equiv |\Phi_1|^2 - |\Phi_2|^2$, which is the Ising nematic order
parameter \cite{nematic,adrian}. This measures fluctuations which break the
orthorhombic symmetry, $C_{4v}$, of the square lattice to tetragonal symmetry,
$C_{2v}$, by making a choice of the SDW order between the ${\bf K}_1$ and
${\bf K}_2$ wavevectors.  This operator is a component of the $({\bf 9}_4,
{\bf 1}_3)$ operators, and to obtain its scaling dimension we computed and
analyzed perturbative series within the MZM (to six loops) and $\overline{\rm
  MS}$ (to five loops) expansion schemes.  Details on the perturbative
calculations and the series will be reported elsewhere.  The comparison of the
analyses within the MZM and $d=3$ $\overline{\rm MS}$ schemes represents a non
trivial check of the results.  In order to estimate the scaling dimension, we
need to resum the series and then to evaluate them at the fixed point values
of the quartic couplings.  The resummation of the series is performed by using
the so-called conformal mapping method, which exploits the knowledege of the
large-order behavior of the expansions~\cite{PRV-01,CPPV-04}.  In this manner
we obtained
\begin{equation}
\mbox{dim}[\lambda_2] = \left\{ 
\begin{array}{ccc} 
1.95(18) -2  &~~& {\rm MZM} \\
1.90(27) -2 &~~& d=3~\overline{\rm MS} \end{array}
\right. \label{dim2}
\end{equation}
These results favor the irrelevance of $\lambda_2$ at the decoupled fixed
point, although the precision of our results does not allow us to state this
conclusively. In any case, it is clear that $\vert \mbox{dim}[\lambda_2]
\vert$ is nearly zero, indicating that $\lambda_2$ hardly flows under the
renormalization group (RG).

Finally, the $\lambda_3, \lambda_3^\prime$ couplings in Eq.~(\ref{lambda}) involve
composite $\Phi$ operators (which are components of $({\bf 6}_4, {\bf 3}_3)$),
which measure spiral spin correlations, as is easily deduced 
from Eq.~(\ref{SPhi}). By an analysis as above, we found
\begin{equation}
\mbox{dim}[\lambda_3,\lambda_3^\prime] = \left\{ 
\begin{array}{ccc} 
1.16(8) -2  &~~& {\rm MZM} \\
1.24(8) -2 &~~& d=3~\overline{\rm MS} \end{array}
\right. ,
\end{equation}
indicating that such perturbations are clearly irrelevant.

We have now established one of the main results of our paper: the only important coupling
of the nodal fermions to the SDW order is $\lambda_2$, which couples the fermionic 
and SDW contributions to the Ising nematic order $\phi$.

Let us now compute of the influence of the SDW fluctuations
on the fermionic spectral functions perturbatively in $\lambda_2$. At leading
order, we use the propagator for $\phi$, implied by its scaling dimension in
Eq.~(\ref{dim2})
\begin{equation}
G_\phi ({\bf p}, \omega)  \sim \left[ v^{\ast 2} {\bf p}^2 +\omega^2 \right]^{-1/2 - {\rm dim}[\lambda_2]} \label{gphi},
\end{equation}
where ${\bf p}$ and $\omega$ are the momentum and imaginary frequency carried by $\phi$. 
To second order in $\lambda_2$, 
for the fermion Green's function $G_\Psi$ we obtain the self energy
\begin{equation}
\Sigma_\Psi ({\bf q}, \Omega) = \lambda_2^2 \int \frac{d^2 p}{4 \pi^2} \frac{d\omega}{2 \pi} 
G_\phi ({\bf p}, \omega) \tau^x G_\Psi ({\bf q} - {\bf p}, \Omega - \omega) \tau^x \label{sig}
\end{equation}
From the finite temperature ($T$) generalization of Eq.~(\ref{sig}), we obtain at the nodal point, the retarded, real frequency self energy
$\mbox{Im} \Sigma_\Psi (0,0) \sim T^{1 - 2 \, {\rm dim}[\lambda_2]}$ in the quantum-critical 
region \cite{qptd}. This is the leading $T$ dependence for $\mbox{dim}[\lambda_2] < 0$. For $\mbox{dim}[\lambda_2] >0$, the RG flow must be computed to higher order in $\lambda_2$, and if 
there is a fixed point with a non-zero $\lambda_2^\ast$, then $T$ is 
characteristic energy scale for quasiparticle damping.
Given the near-marginality
of $\lambda_2$ (Eq.~(\ref{dim2})), it is valid to ignore its RG flow in experimental
applications, and so we obtain
a quasiparticle scattering rate which is practically linear in $T$.
Similarly, at the nodal wavevectors, a nearly linear frequency dependence is found in the fermion self energy
at $T=0$. These are phenomenologically attractive features \cite{seamus2}.

In describing the ${\bf q}$ dependence of the fermion spectrum, we can use results 
from the computation by Kim {\em et al.} \cite{kim}.  They used a model
\cite{qptd} in which the $\phi$ field was the primary order parameter
undergoing phase transition, and not a composite of the SDW order as in our
theory above. In such a primary $\phi$ theory, the $\lambda_2$ coupling is
strongly {\em relevant\/} at the decoupled fixed point \cite{qptd} (in
contrast to the nearly marginal $\lambda_2$ in our theory), and a RG analysis is necessary to find a
fixed point at non-zero $\lambda_2$. No such fixed point was found in the
$3-d$ expansion of Ref.~\onlinecite{qptd} for any number of fermion flavors, while
Kim {\em et al.} \cite{kim} noted that a second order transition was present
$d=2$ in the limit of an infinite number of fermion flavors. Regardless of the
status of this possible fixed point for the physical case, 
for our purposes here, we note that Kim
{\em et al.}'s computation of the fermion spectrum differed from
Eq.~(\ref{sig}) only in their assumed form for $G_\phi$, with Eq.~(\ref{gphi})
replaced by $G_\phi^\prime \sim [(\omega^2 + v_F^2 p_x^2 + v_\Delta^2
p_y^2)^{1/2}$+ $(\omega^2 + v_F^2 p_y^2 + v_\Delta^2 p_x^2)^{1/2}]^{-1}$, the
one-loop contribution of fermion fluctuations to the $\phi$ propagator. Such
contributions are also present in our theory, as an order $\lambda_2^2$
contribution to $G_\phi$ by $G_\phi^{-1} \rightarrow G_\phi^{-1} + \lambda_2^2
G_{\phi}^{\prime -1}$; given that $\lambda_2$ is nearly marginal at the
decoupled fixed point, it is reasonable that this correction, which is
formally higher order in $\lambda_2$, should be included in Eq.~(\ref{sig}).

With $G_\phi$ renormalized with fermion loop contributions as above, our
results for the fermion spectral function have the same qualitative form as
those of Kim~{\em et al.} \cite{kim}. However, it is important to note that
the underlying quantum critical point, and the resulting justification of the
computation, are very different from theirs. In particular, an important
feature of our critical point is that there are no restrictions on the
renormalized values of the fermion velocities, $v_F$ and $v_\Delta$, and there
is no difficulty in them acquiring ratios as large as $v_F/v_\Delta \approx
20$ observed in some cuprates.  As shown by Kim {\em et al.} \cite{kim}, for
large $v_F/v_\Delta$, the fermion excitations are strongly damped by the
$\phi$ fluctuations, except in narrow arc-like regions around the nodal
points; such ``Fermi arcs'' have been noted in a variety of 
experiments \cite{mohit1,mohit2,seamus}.

We can also use our results here for the scaling dimensions of the quadratic perturbations
at the O(4)$\otimes$O(3) fixed point to analyse the influence of quenched disorder. Unless
magnetic impurities are explicitly introduced, the quenched disorder in the cuprates is spin
rotation invariant, and we assume so in our discussion here. 
Such disorder can only couple to operators, $\mathcal{O}_{\bf D}$,  which transforms as
$({\bf D}_4, {\bf 1}_3)$. At quadratic order in $\varphi_{ai}$, the only allowed values are ${\bf D}={\bf 1},{\bf 9}$, and the
scaling dimension of the associated couplings were quoted in Eqs.~(\ref{dim1},\ref{dim2}). 
After an average over disorder in a replica analysis, the perturbation to $\mathcal{L}_\Phi^\ast$
can be written as
\begin{equation}
\gamma_{\bf D} \sum_{\ell,m}
\int d^2 x  d \tau d \tau^\prime \mathcal{O}_{\bf D}^\ell (x, \tau) \mathcal{O}_{\bf D}^m (x, \tau^\prime)
\end{equation}
where $\ell,m$ are replica indices. A computation of scaling dimensions now shows that
the maximum value of $\mbox{dim}[\gamma_{\bf D}]$ is for ${\bf D}={\bf 9}$, and so the most
relevant disorder-induced coupling is $\gamma_{\bf 9}$ with
\begin{equation}
\mbox{dim}[\gamma_{\bf 9}] = 2\, \mbox{dim}[\lambda_2] + 2,
\end{equation}
which is clearly strongly relevant.
One of the components of $\mathcal{O}_{\bf 9}$ is $\phi$, the Ising nematic order, while
the remaining 8 components correspond to the complex charge density wave order parameters
$\Phi_{1}^2$, $\Phi_2^2$, $\Phi_1 \cdot \Phi_2$, $\Phi_1 \cdot \Phi_2^\ast$
at the wavevectors $2 {\bf K}_1$, $2 {\bf K}_2$, ${\bf K}_1 + {\bf K}_2$, ${\bf K}_1 - {\bf K}_2$
respectively. Quenched disorder will therefore induce these orders
in glassy configurations, as was studied numerically in Refs.~\onlinecite{adrian,robertson}.

This paper has shown that the well-established SDW order
of the cuprate superconductor series \cite{jtran,lake,jtran2,boris,mesot} has precursor
quantum fluctuations which can naturally explain key features \cite{mohit1,mohit2,seamus,seamus2} of the spectrum of single electron
excitations. We showed that an Ising nematic order, associated with fluctuations which
reduce the square lattice symmetry to a rectangular symmetry \cite{nematic}, was uniquely
selected as a composite of the SDW order which couples most
efficiently to the low energy single-electron excitations.

We thank Erez Berg, Eduardo Fradkin, Eun-Ah Kim, Steven Kivelson, Michael Lawler, and Cenke Xu for useful discussions.
This research was supported by NSF grant DMR-0537077.


\begin{thebibliography}{99}

\bibitem{jtran}  J.~M.~Tranquada, chapter in 
{\em Treatise of High Temperature Superconductivity\/} by J.~R.~Schrieffer, to be published, arXiv:cond-mat/0512115.

\bibitem{demler} E.~Demler, S.~Sachdev, and Y.~Zhang, Phys. Rev. Lett. {\bf
    87}, 067202 (2001); Y.~Zhang, E.~Demler, and S.~Sachdev, Phys. Rev. B {\bf
    66}, 094501 (2002).

\bibitem{katano} S.~Katano, M.~Sato, K.~Yamada, T.~Suzuki, and T.~Fukase, Phys. Rev.
B {\bf 62}, R14677 (2000).

\bibitem{lake} B.~Lake {\em at al.},
Nature {\bf 415}, 299 (2002).

\bibitem{jtran2} J.~M.~Tranquada, C.~H.~Lee, K.~Yamada, Y.~S.~Lee, L.~P.~Regnault, and
H.~M.~R\o nnow, Phys. Rev. B 69, 174507 (2004).

\bibitem{boris} B.~Khaykovich, S.~Wakimoto, R.~J.~Birgeneau, M.~A.~Kastner, Y.~S.~Lee, 
P.~Smeibidl, P.~Vorderwisch, and K.~Yamada,
Phys. Rev. B {\bf 71}, 220508 (2005).

\bibitem{mesot} J.~Chang, Ch.~Niedermayer, R.~Gilardi, N.~B.~Christensen,
  H.~M.~R\o nnow, D.~F.~McMorrow, M.~Ay, J.~Stahn, O.~Sobolev, A.~Hiess,
  S.~Pailhes, C.~Baines, N.~Momono, M.~Oda, M.~Ido, and J.~Mesot,
  arXiv:0712.2181.
  
\bibitem{zachar} O.~Zachar, S.~A.~Kivelson, and V.~J.~Emery, Phys. Rev. B {\bf
    57}, 1422 (1998).
  
\bibitem{vicari1} M.~De Prato, A.~Pelissetto, and E.~Vicari Phys. Rev. B {\bf
    74} 144507 (2006).
  
\bibitem{nematic} S. A. Kivelson, E. Fradkin, and V. J. Emery, Nature {\bf
    393}, 550 (1998).
  
\bibitem{kim} E.-A.~Kim, M.~J.~Lawler, P.~Oreto, E.~Fradkin, and
  S.~A.~Kivelson, arXiv:0705.4099.

\bibitem{qptd} M.~Vojta, Y.~Zhang, and S.~Sachdev,
Phys. Rev. Lett. {\bf 85}, 4940  (2000).


\bibitem{frey} E.~Frey and L.~Balents,
Phys. Rev. B {\bf 55}, 1050 (1997).

\bibitem{berg} E.~Berg, C.-C.~Chen, and S.~A.~Kivelson,
Phys. Rev. Lett. {\bf 100}, 027003 (2008).

\bibitem{adrian} A.~Del Maestro, B.~Rosenow, and S.~Sachdev, Phys. Rev. B {\bf 74}, 024520 (2006).

\bibitem{PRV-01}
A.~Pelissetto, P.~Rossi, and E.~Vicari,
Phys. Rev. B {\bf 63}, 140414(R) (2001).

\bibitem{CPPV-04}
P.~Calabrese, P.~Parruccini, A.~Pelissetto, and E. Vicari,
Phys. Rev. B {\bf 70}, 174439 (2004).

\bibitem{seamus2} J.~W.~Alldredge {\em et al.\/}, arXiv:0801.0087.

\bibitem{mohit1} M.~R.~Norman {\em et al.\/},
Nature {\bf 392}, 157-160 (1998).

\bibitem{mohit2} A.~Kanigel {\em et al.\/},
Nature Physics {\bf 2}, 447-451 (2006).

\bibitem{seamus} J.~Lee, J.~A.~Slezak, and J.~C.~Davis,
J. Phys. Chem. Solids {\bf 66}, (2005).

\bibitem{robertson} J.~A.~Robertson, S.~A.~Kivelson, E.~Fradkin, A.~C.~Fang, 
and A.~Kapitulnik, Phys. Rev. B {\bf 74}, 134507 (2006).
\end{thebibliography}
\end{document}